\documentclass[12pt]{article}
\pdfoutput=1
\usepackage{amsmath}
\usepackage{amsfonts}
\usepackage{amssymb}
\usepackage{graphicx, rotating}
\usepackage{epsfig}
\usepackage{latexsym}
\usepackage{graphicx}
\usepackage{color}
\usepackage{amsmath,bm,amssymb}

\usepackage[english]{babel}
\usepackage[colorlinks,allcolors=blue]{hyperref}
\usepackage[numbers,sort&compress]{natbib}
\usepackage[normalem]{ulem}

\newcommand{\Slash}[1]{{\ooalign{\hfil#1\hfil\crcr\raise.167ex\hbox{/}}}}

\newcommand{\beq}{\begin{equation}}  \newcommand{\eeq}{\end{equation}}
\newcommand{\bef}{\begin{figure}}  \newcommand{\eef}{\end{figure}}
\newcommand{\bec}{\begin{center}}  \newcommand{\eec}{\end{center}}
\newcommand{\non}{\nonumber}  
\newcommand{\laq}[1]{\label{eq:#1}}  

\newcommand{\Eq}[1]{Eq.\,(\ref{eq:#1})}

\newcommand{\Sec}[1]{Sec.\,\ref{chap:#1}}

\newcommand{\vev}[1]{ \left\langle {#1} \right\rangle }

\newcommand{\lac}[1]{\label{chap:#1}}
\newcommand{\U}[1]{{\rm U{#1}} }
\newcommand{\SU}[1]{{\rm SU{#1} } }

\def\({\left(}
\def\){\right)}

\def\O{\mathcal{O}}

\newcommand{\AND}{~{\rm and}~}

\newcommand{\GEV}{ {\rm ~GeV} }
\newcommand{\TEV}{ {\rm ~TeV} }

\def\a{\alpha}
\def\b{\beta}

\def\d{\delta}

\def\g{\gamma}

\def\m{\mu}

\def\x{\xi}

\def\D{\Delta}

\def\F{\Phi}

\def\tl{\tilde}
\def\*{\dagger}

\usepackage[affil-it]{authblk}
\usepackage[left=2cm,right=2cm,top=2cm,bottom=2cm,a4paper]{geometry}

\begin{document}
\begin{titlepage}
\begin{center}
%\allowdisplaybreaks
\setcounter{footnote}{0}
\setcounter{figure}{0}
\setcounter{table}{0}

\hfill   TU-1XXX\\
\vspace{3.5cm}

{\Large\bf 
A Novel Probe of Supersymmetry in Light of Nanohertz Gravitational Waves}

\vskip .75in

{ \large Kai Murai~ and~ Wen Yin}

\vskip 0.25in

\begin{tabular}{cc}
& {\em Department of Physics, Tohoku University, }\\
& {\em Sendai, Miyagi 980-8578, Japan}\\[.3em]

\vspace{12pt}
\vspace{1.5cm}

\end{tabular}

%%%%%%%%%%%%%%%%%%%%%%%%%%%%%%%%%%%%%%%%%%%
\begin{abstract}
A new era of exploring the early Universe may have begun with the recent strong evidence for the stochastic gravitational wave (GW) background from the data reported by NANOGrav, EPTA, PPTA, and CPTA.
Inspired by this, we propose a new potential source of stochastic GWs in the minimal supersymmetric standard model (MSSM), which could be the theory at a very high energy scale.
This source is the ``axion" field in the Higgs multiplets when the Higgs field takes a large value along the D-flat direction in the early Universe, for example, during inflation.
The axion motion triggers the instability of the standard model ${\rm U}(1)$ and/or ${\rm SU}(3)$ gauge fields, producing stochastic GWs during the inflation.
This scenario can be seen as a simple UV completion of the commonly studied models where an axion spectator/inflaton is coupled to a hidden ${\rm U}(1)$ or ${\rm SU}(N)$ gauge field without matter fields.
Thus the nanohertz GWs may be a sign of supersymmetry.
Primordial magnetic field production is also argued. In addition, we point out the simple possibility that this axion within the MSSM drives inflation.

\noindent
\end{abstract}
%%%%%%%%%%%%%%%%%%%%%%%%%%%%%%%%%%%%%%%%%%%

\end{center}
\end{titlepage}
\setcounter{footnote}{0}
\setcounter{page}{1}

%%%%%%%%%%%%%%%%%%%%%%%%%%%%%%%%%%%%%%%%%%%
\section{Introduction}
%%%%%%%%%%%%%%%%%%%%%%%%%%%%%%%%%%%%%%%%%%%

The groups NANOGrav, EPTA, PPTA, and CPTA have recently reported data suggesting the presence of a stochastic gravitational wave (GW) background in the frequency band of ${\cal O}(1\,\text{--}\,10)$\,nHz~\cite{NANOGrav:2023gor,Antoniadis:2023ott,Reardon:2023gzh,Xu:2023wog}. In this frequency range, supermassive black holes emerge as a leading candidate for the origin of these GWs. Additionally, other new physics interpretations are plausible (see also Refs.~\cite{Kitajima:2023cek,Bai:2023cqj,Ellis:2023tsl,Franciolini:2023pbf,Fujikura:2023lkn,Guo:2023hyp,Han:2023olf,Li:2023yaj,Megias:2023kiy,Vagnozzi:2023lwo,Yang:2023aak,Zu:2023olm} for the potential source from beyond Standard Model (SM) proposed after the announcement). A new era of exploring the early Universe may have begun. In this paper, at the same time to introduce a new explanation of the data, we will point out that searching for the stochastic GWs in general frequency can be useful to probe supersymmetry (SUSY) in the inflationary Universe. 

The stochastic GWs can be generated during inflation.
Cosmic inflation~\cite{Starobinsky:1980te,Guth:1980zm,Sato:1980yn,Linde:1981mu} provides the origin of the scalar perturbations that seed the anisotropies observed in the cosmic microwave background (CMB) and the large-scale structures.
Moreover, quantum fluctuations of the tensor components of the spacetime metric predict the existence of the stochastic primordial GWs.
However, the GW spectrum is almost scale-invariant and limited from above on the CMB scales.
Thus, it is unlikely that such GWs explain the stochastic GWs reported by the PTA experiments.
If some fields other than the inflaton shows nontrivial dynamics during inflation, scale-dependent GWs can be generated.

One possibility is $\U(1)$ gauge fields.
For example, a $\U(1)$ gauge field can {couple} to the inflaton through a topological coupling $\propto \phi F_{\mu \nu} \tilde{F}^{\mu \nu}$. 
In this case, the gauge fields exponentially grow sourced by the inflaton motion, while the backreaction gives friction to the inflaton, which realizes slow-roll inflation even with a steep inflaton potential~\cite{Anber:2009ua}.
Such an enhancement of the gauge field was originally studied in the context of primordial magnetogenesis~\cite{Turner:1987bw}.
The enhanced $\U(1)$ gauge field can induce sizable gravitational waves with non-standard properties~\cite{Sorbo:2011rz,Barnaby:2011qe,Anber:2012du,Domcke:2016bkh,Peloso:2016gqs,Adshead:2019igv,Watanabe:2020ctz}.
Moreover, if the $\U(1)$ gauge field couples to a spectator axion that rolls down the potential during inflation, the GWs can also be generated by the same mechanism~\cite{Barnaby:2012xt,Cook:2013xea,Shiraishi:2013kxa,Mukohyama:2014gba,Namba:2015gja,Peloso:2016gqs,Shiraishi:2016yun,Ozsoy:2017blg,Ozsoy:2020ccy,Ozsoy:2021onx}.
Note that the amplified gauge field also sources the scalar perturbations, which can be highly non-Gaussian~\cite{Barnaby:2010vf,Barnaby:2011vw,Barnaby:2011qe,Anber:2012du,Meerburg:2012id,Barnaby:2012xt,Linde:2012bt,Ferreira:2014zia,Bartolo:2015dga,Ozsoy:2017blg,Almeida:2019hhx} and/or large enough to produce primordial black holes~\cite{Linde:2012bt,Bugaev:2013fya,Erfani:2015rqv,Garcia-Bellido:2016dkw,Domcke:2017fix,Cheng:2018yyr,Ozsoy:2020kat} and provide a probe of the scenario.

Another possibility is non-Abelian gauge fields.
The model in which the inflaton couples to $\SU(2)$ gauge fields through the topological coupling $\propto \phi G_{\mu \nu}^a \tilde{G}_a^{\mu \nu}$, Chromo-Natural Inflation~\cite{Adshead:2012kp}, has attracted much attention.
In this model, the inflaton motion induces a homogeneous, isotropic, and attractor solution of the $\SU(2)$ gauge fields~\cite{Maleknejad:2013npa,Domcke:2018rvv,Wolfson:2020fqz,Wolfson:2021fya}, while the gauge field background backreacts to the inflaton and slows down the inflaton motion.
Under the existence of this gauge field background, a part of the gauge field perturbations experiences tachyonic instabilities and linearly sources chiral GWs~\cite{Dimastrogiovanni:2012ew,Maleknejad:2016qjz}.
Due to the overproduction of the GWs, the original chromo-natural inflation scenario has been excluded by the CMB observations~\cite{Adshead:2013qp,Adshead:2013nka}.
However, there are some models to exploit this mechanism to generate the chiral GWs without violating the observational constraints.
If the homogeneous gauge field arises after the CMB scale exits the horizon during inflation, the observable GWs can be generated on smaller scales~\cite{Obata:2014loa,Obata:2016tmo,Domcke:2018rvv,Fujita:2022jkc}.
Moreover, if the pseudoscalar field coupled to the gauge fields works as a spectator field, the dynamics of the pseudoscalar and gauge fields is not responsible for the scalar perturbations, and then the observable chiral GWs can be generated without spoiling the success of inflation~\cite{Dimastrogiovanni:2016fuu,Ishiwata:2021yne}.
In this case, the contribution of the gauge field perturbations can dominate the total primordial GWs, resulting in chiral and non-Gaussian GWs~\cite{Dimastrogiovanni:2016fuu,Agrawal:2017awz,Thorne:2017jft,Agrawal:2018mrg,Dimastrogiovanni:2018xnn,Fujita:2018vmv,Fujita:2018ndp,Fujita:2021flu}.
Although these studies are conducted for the $\SU(2)$ gauge fields, the $\SU(N)$ gauge fields coupled to the pseudoscalar also have homogeneous and isotropic solutions corresponding to $\SU(2)$ subgroups in $\SU(N)$~\cite{Fujita:2021eue}.
At the linear level, the behavior of the perturbations is also analogous to the $\SU(2)$ case, and then the GW generation is expected~\cite{Fujita:2022fff}.
The isotropization of the gauge field configuration is dynamically confirmed in the numerical simulations~\cite{Murata:2022qzz}.

Note that, in the above scenarios of the GW generation, it is typically assumed that the gauge fields coupled to the pseudoscalar are different from the ones of the SM.
Generating GWs using {the SM gauge groups $\U(1)$, $\SU(2)$, and $\SU(3)$ is challenging or at least is not well understood}.
{This is primarily because the SM includes charged particles under these gauge groups. The Schwinger effect subsequently generates these charged fields, providing a backreaction that inhibits the growth of the gauge fields. 
In the case of $\U(1)$, the upper bound of the produced gauge fields is derived by taking into account this effect~\cite{Domcke:2018eki}. 
The Schwinger effect in the $\SU(2)$ case is also studied in Refs.~\cite{Lozanov:2018kpk,Domcke:2018gfr,Mirzagholi:2019jeb,Adshead:2022ecl}. In particular, the effect is small if the gauge coupling is small and the number of light charged particles is small.
However, it is still not clear if the condition is satisfied for the SM $\SU(2)$ and, especially, $\SU(3)$.}
In contrast, in our proposed scenario, we plan to utilize the SM's $\U(1)$ and $\SU(3)$ gauge groups, where the Higgs field {is assumed to have} 
large values during inflation.%
\footnote{See Refs.\,\cite{Dvali:1995ce,Banks:1996ea,Choi:1996fs,Jeong:2013xta,Co:2018phi,Ho:2019ayl, Matsui:2020wfx}
for the discussions of large Higgs field during inflation leading to a stronger QCD to make the QCD axion heavier during the inflation in order to reduce the QCD axion abundance or the isocurvature perturbation.
This is particulary efficient in the stochastic axion scenarios~\cite{Graham:2018jyp,Takahashi:2018tdu, Ho:2019ayl, Matsui:2020wfx} (See also Refs.~\cite{Alonso-Alvarez:2019ixv, Nakagawa:2020eeg,  Marsh:2019bjr, Kitano:2021fdl, Murai:2023xjn} for applications.)} 
This effectively imparts significant mass to the charged particles that are otherwise undesirable, effectively decoupling them from the system. We will demonstrate that this setup naturally occurs in the minimal SUSY extension of the SM (MSSM).

The MSSM, one of the leading extensions of physics beyond the SM, is compatible with the grand unified theory (GUT), alleviates the naturalness problem, and provides a candidate for WIMP dark matter. The SUSY scale might exceed the detection range of recent accelerator experiments, which have excluded a large parameter region of the low-scale SUSY scenarios. However, high-scale SUSY mitigates the challenges of proton decay, SUSY flavor, SUSY CP, and gravitino problems, inherent to low-scale SUSY (see Refs.~\cite{ArkaniHamed:2004yi,Giudice:2004tc, Ibe:2006de,Ibe:2011aa,ArkaniHamed:2012gw,Yamaguchi:2016oqz,Yin:2016shg, Yanagida:2016kag, Yanagida:2018eho,Yanagida:2019evh} for various high-scale SUSY scenarios), and thus is very consistent to the experimental results as well as cosmology.  It is particularly important to develop ideas or methods for probing this. In the MSSM, there are multiple almost flat directions of the scalar potential, especially the so-called D-flat direction. Therefore, in the early Universe, such as during inflation, it is natural for the Higgs field values to deviate significantly from the electroweak scale.

In this paper, we point out that if the Higgs fields in the early Universe have much larger field values than the SUSY scale, a pseudo Nambu-Goldstone (NG) boson, or an axion, emerges due to the spontaneous breaking of the approximate ``Peccei-Quinn" (PQ) symmetry. This bears similarities to the original PQ-Weinberg-Wilczek (PQWW) model~\cite{Peccei:1977hh, Peccei:1977ur, Weinberg:1977ma,Wilczek:1977pj}, although our axion description is a valid approximation only in the early Universe, thereby circumventing constraints observed in the present Universe. With the matter fields also being heavy, the axion couples to the unbroken photon and gluon via the anomaly of the PQ symmetry. This realization of the axion coupling provides a simple UV realization of the setup described earlier.

Even if the SUSY scale is high, the GWs can be produced, and this scenario not only offers a natural and minimal UV realization of earlier work on the GW generation by gauge fields, but also gives a new opportunity to probe high-scale SUSY. Indeed, the nanohertz GW signal could potentially serve as a signature of SUSY. Moreover, the axion could drive inflation, thereby proposing one of the minimal SUSY models for this phenomenon.

This paper is organized as follows. 
In \Sec{setup}, we discuss a generic idea for having the GW generation by using the SM photon and gluon field and discuss the GW spectrum. 
In \Sec{MSSM}, we use the MSSM to UV complete the setup. 
The last section is devoted to the discussion and conclusions, where we also discuss the generation of the primordial magnetic field.

%%%%%%%%%%%%%%%%%%%%%%%%%%%%%%%%%%%%%%%%%%%
\section{Mechanism in nutshell and GWs} 
\lac{setup}
%%%%%%%%%%%%%%%%%%%%%%%%%%%%%%%%%%%%%%%%%%%

To illustrate our idea, let us consider an effective Lagrangian during inflation, in addition to the SM one,
\beq\laq{setup}
\d {\cal L} = \frac{ a }{8\pi f_a} (c_\gamma \a F \tilde{F} +c_g \a_s G^a\tl G_a +\cdots)
.
\eeq
Here, $F \AND G$ ($\tl F, \tl G$) represent the photon and gluon field strengths (and their duals), with $\a \AND \a_s$ being their corresponding coupling constants, and $\cdots$ denote the other components of $\U(1)\times \SU(2)$ field strengths and couplings. $a$ represents the axion (or a similar particle). $c_\g$ and $c_g$ denote the axion-photon and axion-gluon coupling coefficients, respectively, and $f_a$ is the decay constant of $a$.
As we have mentioned earlier, this setup might not efficiently generate GWs since light charged particles can interfere with the tachyonic production of gauge fields.

To circumvent this issue, we introduce the Hubble-induced mass to the Higgs field
\beq
\d V=- \x H_{\rm inf}^2 |\F|^2 ,
\eeq
where $\F$ represents the SM Higgs doublet, and $H_{\rm inf}$ signifies the Hubble parameter during inflation. 
$\x$ is a coupling, which is related to the Higgs non-minimal coupling or Higgs direct coupling to the inflaton. We consider $\x=\O(1)>0$, leading to a scenario where the Higgs field acquires a negative mass squared during inflation that is more efficient than the Hubble friction. Consequently, the Higgs field is driven to a position far from the vacuum expectation value (VEV) of the electroweak scale. The field value during inflation may be determined by the balance between the Hubble-induced term and another term, e.g., a quartic term or a higher dimensional term.  For later convenience, we assume that the norm of $\F$ during inflation is a free parameter $|\F|=h_{\rm inf}$.%
\footnote{We do not discuss the Higgs potential instability since this will be solved by SUSY in the next section.}
This gives masses to the matter fields. We consider that neutrinos are also heavy\footnote{This is for simplicity of discussion. Even if we consider such a model that the neutrinos remain light, our conclusions do not change since they do not induce the Schwinger effect for preventing the growth of the $\U(1), \SU(3)$ gauge fields. } by assuming that the mass is generated via higher dimensional terms, $\F L \F L$, with $L$ being the lepton doublet or that it is generated via the see-saw mechanism~\cite{Yanagida:1979as,Glashow:1980,Gell-Mann:1979vob, Minkowski:1977sc,Mohapatra:1979ia}. Indeed, the W and Z- bosons become as heavy as $h_{\rm inf}.$ The mass of the lightest charged particle, i.e., the electron, is given by $m_{\rm lightest}\sim 10^{13}\GEV \frac{h_{\rm inf}}{M_{\rm pl}}$. We also note that the lightest charged particle mass depends on the UV completion and it may be even heavier. This can be understood by considering a higher dimensional term, $\frac{|\F|^2}{M^2}\F^* L e$ with $M$ being the energy scale of the more fundamental physics, and $e$ representing the right-handed lepton.

In the effective {field} theory (EFT) by integrating out the massive fields, i.e., the renormalization scale below $m_{\rm lightest},$ 
 we only have the fields of
 \beq
\text{photons~}F,~ \text{gluons~}G^a,\text{~and}~ a.
 \eeq
We also comment that the potential for the axion, \beq V[a],\eeq should be added  to the above Lagrangian. Depending on the potential shape of $V[a]$, the axion can be the inflaton itself or a spectator field. 

In this EFT, depending on the choice of $c_\g, c_g$, we may have different kinds of the GW production.
{For simplicity, in the following,} we assume that the strengths of these interactions are hierarchical and discuss the GW generation when the system is approximately composed of the axion and one of the gauge fields.
{Note that, however, if the two interactions have comparable couplings, the two mechanisms can simultaneously work, which {should} result in the distinctive GW spectrum, although we cannot find an existing work.}
In both setups, the gauge field is amplified by the motion of the axion, which is controlled by the axion velocity.
Conventionally, we parameterize the axion {velocity} 
by
\begin{align}
    \xi_a
    =
    \frac{c_\gamma \alpha}{4 \pi f_a H_\mathrm{inf}}\dot{a}
    \quad \mathrm{or} \quad 
    \frac{c_g \alpha_s}{4 \pi f_a H_\mathrm{inf}}\dot{a}
    \ .
\end{align}
{As we will discuss shortly, for efficient GW production, the motion needs to satisfy $|\dot{a}|\gtrsim \O(100-1000) c_{\g,g}^{-1} H_{\rm inf} f_a$. This implies that the (height of the axion potential)$/f_a^2$ should be larger than $H^2_{\rm inf}$ for the motion to be fast enough.}

%%%%%%%%%%%%%%%%%%%%%%%%%%%%%%%%%%%%%%%%%%%
\paragraph{{Axion and $\U(1)$ gauge field}}
%%%%%%%%%%%%%%%%%%%%%%%%%%%%%%%%%%%%%%%%%%%
First, we consider the limit of the small axion-gluon coupling.
For a sufficiently low inflation scale, we have essentially the same setup for generating the GWs by an axion-hidden $\U(1)$ setup. 
It is noteworthy that the charged particles are decoupled and the $\U(1)$ field does not suffer from the Schwinger effect preventing the exponential growth of the gauge fields, which is an important feature of our setup. 
Through the coupling to the rolling axion, the perturbations of the gauge field are exponentially amplified, which in turn sources the scalar and tensor perturbations at the second-order level.
While the observable signal on the scales smaller than the CMB scale is expected for $|\xi_a| \gtrsim 5$, the backreaction is non-negligible for $|\xi_a| \gtrsim 4.7$~\cite{Peloso:2016gqs}.
In fact, the axion velocity oscillates due to the backreaction as confirmed in the numerical simulation~\cite{Cheng:2015oqa,Notari:2016npn,Sobol:2019xls,DallAgata:2019yrr,Domcke:2020zez,Gorbar:2021rlt,Caravano:2022epk,Durrer:2023rhc,Garcia-Bellido:2023ser,Figueroa:2023oxc}.
Such an oscillating behavior of the axion velocity is imprinted into the GW spectrum, which can be probed by the PTA and interferometer experiments~\cite{Garcia-Bellido:2023ser}.%  
\footnote{The analysis in Ref.~\cite{Garcia-Bellido:2023ser} assumes homogeneous inflaton and considers only homogeneous backreaction from the gauge field. 
However, it is recently suggested that the inhomogeneity of the inflaton and backreaction can affect the evaluation of the GWs~\cite{Figueroa:2023oxc}.}

{
In contrast, in scenarios involving charged fields, a significant backreaction occurs when $|\xi_a| \gtrsim 4$, which prevents the exponential growth \cite{Domcke:2018eki,Domcke:2019qmm,Fujita:2022fwc}. However, in our setup, this suppression effect is absent. }

%%%%%%%%%%%%%%%%%%%%%%%%%%%%%%%%%%%%%%%%%%%
\paragraph{{Axion and $\SU(2)$ gauge field}}
%%%%%%%%%%%%%%%%%%%%%%%%%%%%%%%%%%%%%%%%%%%

In {the other} %another 
limit, we have the axion-pure $\SU(N)$ system with $N=3$.
In this case, it is noted that the system of rolling axion and $\SU(2)_{\rm sub}$ subgroup can have an attractor solution with nonzero gauge field strength~\cite{Fujita:2021eue}.%
\footnote{It may be important to study the Schwinger effect by the charged vector fields corresponding to $\SU(N)/\SU(2)_{\rm sub}$.} 
{%
In this case, the analysis of the $\SU(2)$ model can be applied to the $\SU(3)$ case for the linear perturbations~\cite{Fujita:2022fff}.
If the axion velocity is sufficiently large, $|\xi_a| > 2$, the gauge field has an isotropic background solution, which is expected to be realized for larger $|\xi_a|$~\cite{Domcke:2018rvv}.
Depending on the axion potential, such an emergence of the gauge field background occurs after the horizon exit of the CMB scale.
Since the GWs are mainly amplified around the Hubble scale, an enhancement of the GW spectrum is expected on smaller scales.
For example, a model where the $\SU(2)$ gauge field acquire nonzero background in the course of inflation predicts the generation of the GWs that can be probed in the future GW interferometers~\cite{Fujita:2022jkc}.%
\footnote{The possibility of the nanohertz GWs is pointed out in the delayed emergence of the gauge field in different setups~\cite{Obata:2014loa,Obata:2016tmo}.}
Note that, although the larger amplitude of the GWs will be obtained for the steeper axion potential, the backreaction from the gauge field becomes non-negligible at some point, and the stationary dynamics of the axion and gauge field becomes unstable~\cite{Ishiwata:2021yne}.
}

\begin{figure}[!t]
    \begin{center}  
         \includegraphics[width=145mm]{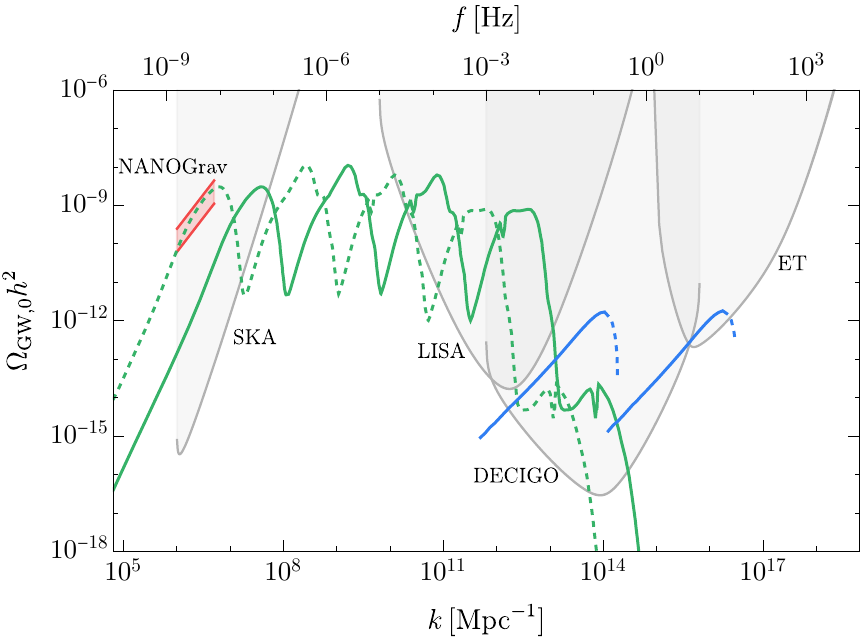}
    \end{center}
    \caption{%
        The GW spectrum predicted in the scenario with axion and gauge fields.
        The green solid line shows the GW spectrum in the $\U(1)$ scenario in Ref.~\cite{Garcia-Bellido:2023ser}, and the green dashed line is shifted in the frequency direction.
        The blue lines show those in the $\SU(2)$ scenario with different model parameters in Ref.~\cite{Fujita:2022jkc}.
        The blue dashed lines represent the region where the backreaction of the gauge field may affect the GW spectrum.
        The red-shaded region corresponds to the 2$\sigma$ range of the NANOGrav 15\,yr dataset~\cite{NANOGrav:2023gor} with the best-fit power, $\Omega_{\mathrm{GW},0}h^2 \propto k^{1.8}$ ($\gamma = 3.2$ in their notation).
        The gray-shaded regions are the future sensitivities of the Square Kilometre Array (SKA)~\cite{Janssen:2014dka,Weltman:2018zrl}, Laser Interferometer Space Antenna (LISA)~\cite{LISA:2017pwj}, Deci-Hertz Interferometer Gravitational-Wave Observatory (DECIGO)~\cite{Kawamura:2011zz}, and Einstein Telescope (ET)~\cite{Punturo:2010zz}.
        The sensitivity curves are taken from Ref.~\cite{Schmitz:2020syl}.
    }
    \label{fig: GW spectrum} 
\end{figure}

As an example, we show the GW spectra predicted in scenarios in which the axion is coupled to $\U(1)$~\cite{Garcia-Bellido:2023ser} or $\SU(2)$ gauge fields~\cite{Fujita:2022jkc} in Fig.~\ref{fig: GW spectrum}.
We also show the GW region suggested by the NANOGrav 15\,yr dataset~\cite{NANOGrav:2023gor} and the sensitivities of the future experiments taken from Ref.~\cite{Schmitz:2020syl}.
Although the GW spectrum is numerically obtained assuming a specific shape of the axion potential in Ref.~\cite{Garcia-Bellido:2023ser}, we naively expect that the frequency dependence of the GWs can be shifted by modifying the potential shape or changing the duration of the inflationary epoch.\footnote{{For a constant $\x_a$, it was analytically understood that the GW spectrum is almost scale invaraint~\cite{Barnaby:2011vw}. The amplitude is only relevant to $H_{\rm inf}$ and $\x_a$. }}
Following this naive expectation, we also show a shifted GW spectrum by a green dashed line, which matches the NANOGrav result. {Indeed, the typical frequency of the GWs in each case depends on the e-fold when $|\x_a|$ becomes sizable and thus model dependent.}

%%%%%%%%%%%%%%%%%%%%%%%%%%%%%%%%%%%%%%
%%%%%%%%%%%%%%%%%%%%%%%%%%%%%%%%%%%%%%

%%%%%%%%%%%%%%%%%%%%%%%%%%%%%%%%%%%%%%%%%%%
\section{A novel GW probe of SUSY}
\lac{MSSM}
%%%%%%%%%%%%%%%%%%%%%%%%%%%%%%%%%%%%%%%%%%%
Here, we demonstrate that the above setup, including the axion, can be naturally UV completed in the MSSM. This is possible because (i) the Higgs potential possesses an almost D-flat direction, and (ii) when the Higgs field values are large, the PQ symmetry, which is explicitly broken in the current vacuum, is precise and undergoes spontaneous breaking. Thus an axion appears as in the PQWW model. We will elaborate on the spectrum, which is summarized in Table.~\ref{tab:my_label}, in greater detail. As we will illustrate, in the EFT, by integrating out particles above the mass of $H_{\rm inf}$, we obtain the model in \Eq{setup}, by identifying $A$ to be $a$.

%%%%%%%%%%%%%%%%%%%%%%%%%%%%%%%%%%%%%%%%%%%
\begin{table}[t!]
    \centering
    \begin{tabular}{|c|c|}
        \hline
        Particle, Hubble&  Mass scale  \\
        \hline
        $  Q, u,d,L,e$ & $ M$   \\
        $\tl H^+, \tl H^0, \tl W, \tl Z$ &  $ M$ \\
        $  \tl Q, \tl u,\tl d,\tl L,\tl e$ &  $M$\\
        $ H^+, H^0,  W, Z$ &  $ M$   \\
        \hline
        $ h, \tl{A}$ & $\frac{|F_Z|}{M} $ \\
        $H_{\rm inf}$ & $\frac{F_Z}{M_{\rm pl}}$\\ 
        \hline
        $A $ & $\sqrt{\frac{|F_Z\mu|}{M}} $\\ 
        photon, gluon  & massless\\ 
        \hline
    \end{tabular}
    \caption{%
     Spectrum at the tree-level for the particle contents of our model during inflation with large Higgs values $\sim M$. From top to bottom, the mass scale decreases. The top block denotes the particle with non-vanishing masses at the SUSY limit, the middle is generated by the effects of SUSY breaking and the last block represents the light components due to the PQ symmetry and the gauge symmetry. Here $Q,u,d,L,e$ denote the chiral fermion of the left-handed quark, right-handed up-type quark, right-handed down-type quark, left-handed lepton, and right-handed lepton, respectively. Other fields and parameters are defined in the main text. $\tl X$ denotes the SUSY partner of $X$. $M_{\rm pl}\simeq 2.4\times 10^{18}\GEV$ is the reduced Planck scale.
    }
    \label{tab:my_label}
\end{table}
%%%%%%%%%%%%%%%%%%%%%%%%%%%%%%%%%%%%%%%%%%%
\subsection{D-flat direction and PQ symmetry}

To demonstrate our claim more precisely, we consider the scalar potential of the MSSM. For the sake of clarity and our specific purpose, we consider that the squarks and sleptons have vanishing field values. Thus, the potential is solely composed of the Higgs fields:
\begin{align}
\laq{pot}
V=&(|\mu|^2 +m_{Hu}^2)|H_u|^2+(|\mu|^2 +m_{Hd}^2)|H_d|^2+B\mu (H_u^+ H_d^- + H_u^0 H_d^0)+h.c. \non \\
+ &\frac{g^2}{2} \left|H_u^+ (H_d^0)^*-H_u^0 (H_d^-)^*\right|^2 +\frac{g^2+g'^2}{8}(|H_u|^2-|H_d|^2)^2.
\end{align}
Here, $H_u$ and $H_d$ are the up-type and down-type Higgs doublet fields, respectively, and $|H_u|^2$ and $|H_d|^2$ denote the norm of the doublets. In the first line, we have the $\mu$-term SUSY contribution and the SUSY-breaking soft terms, $m_{Hu}^2$, $m_{Hd}^2$, and $B\mu$. In the second line, we display the D-term contributions, where $g$ and $g'$ represent the $\SU(2)$ and $\U(1)$ couplings, respectively.
 
% the scalar components of the quark and lepton multiplets have the vanishing field value. 

We observe that the D-term exhibits a flat direction: 
\begin{equation}
\text{Almost flat direction: } H_u^0=H_d^0=v,~~ H_u^+= (H_d^-)^*=0.
\end{equation}
In this equation, we have redefined the field so that the field value $v$ is real. By considering the following condition:
\begin{equation}
\laq{largevev}v\gg |m_{Hu,Hd}|, |B \mu|, |\mu|,
\end{equation}
which we will justify shortly, we can neglect the contributions from the first line of the potential. Along the D-flat direction, the potential is zero, and SUSY remains well preserved.

From \Eq{pot},  it is straightforward to ascertain that the charged Higgs $H^+=(H_u^+-(H_d^-)^*)/\sqrt{2}$
gains a mass squared of $M_W^2\equiv \frac{g^2 v^2}{2},$ and and a neutral CP-even Higgs $H^0 = \Re (H_u^0-H_d^0)$ acquires a mass squared of $M_Z^2\equiv \frac{g^2+g'^2}{2}v^2.$
The other components, $h^0= \Re (H_u^0+H_d^0)$, $A^0=\Im (H_u^0+H_d^0)$, $\pi^0=\Im (H_u^0-H_d^0)$, \AND $\pi^+=\frac{1}{\sqrt{2}} (H_u^++(H_d^-)^*)$, remain massless. Here, $\pi^+ \AND \pi^0$ are would-be NG bosons that are eaten by $W^+$ and $Z$ bosons, who gain the masses of $M_W \AND M_Z$, respectively. 

Importantly, $h^0 \AND A^0$ also remain massless. This is because they form an NG multiplet due to the spontaneous breaking of the PQ symmetry, under which each of $H_u$ and $H_d$ carries a charge of 1/2. This symmetry manifests itself when we neglect the terms in the first line of \Eq{pot}.  

Given that we are neglecting SUSY breaking effects for the simplicity of the discussion at this stage, we can easily estimate the masses of the superpartners according to the SUSY relation. For example, the masses of the Higgsinos, denoted $\tl{H^0}$ and $\tl H^+$, are $M_Z$ and $M_W$, respectively. The gauginos, represented as $\tl Z$ and $\tl W$, have the masses of $M_W$ and $M_Z$ respectively. $H^+,\tl H^+, W^+, \AND \tl W^+$ ($H^0,\tl H^0, Z, \AND \tl Z$)  have the same mass because they form a massive vector multiplet.
The particles $\tl h$ and $\tl A$ are massless at this limit and will acquire masses due to the SUSY breaking effects.

% We also have  e.g. massless axino. 

Before ending this part, it is worth emphasizing once again that the discussion thus far hinges on the assumption outlined in \Eq{largevev}, which is aimed at discussing the inflationary spectrum subsequently. This deviates from the MSSM in the present Universe, where the $B\mu$ and $\mu$-terms are comparable to or larger than the  Higgs VEVs. In this case, the PQ  symmetry is badly  broken explicitly, and treating $A$ as the NG boson is not an appropriate description.

\subsection{SUSY breaking and large Higgs during inflation}

In the early Universe, especially during inflation, the Higgs fields can acquire large field values along the D-flat direction. 
With the large field, the PQ symmetry-breaking effect can be neglected in a large class of models.
One can consider the correction to the K\"{a}hler potential of the form 
\beq
\laq{kahler}
\D K \supset \frac{|Z|^2}{M^2} (c_u |H_u|^2+c_d |H_d|^2+ \O(M^{-1})). 
\eeq
Here a chiral multiplet $Z$ acquires an $F$-term, $F_Z$, during inflation. 
We assume that $|F_Z|^2$ is the order of the inflationary potential energy, and $H_{\rm inf}\sim |F_Z|/M_{\rm pl}$ (see \Sec{inf} for a model building).
$M$ is the cutoff scale of the MSSM, and $c_{u,d}$ are $\mathcal{O}(1)$ coefficients. 
The couplings  $Z |H_{u,d}|^2$ have been eliminated through an appropriate redefinition of the fields $H_u$ and $H_d$.

A key assumption is that the PQ symmetry is sufficiently preserved, allowing us to neglect the $H_u H_d$ term in this potential.
More precisely, we anticipate an explicit breaking term of the form
\beq \laq{super} \D W\supset c_1 \frac{Z}{M} \mu H_u H_d+c_2 \frac{Z}{M}\frac{\mu}{M^2} (H_u H_d)^2 + \cdots ,\eeq 
in, e.g., the superpotential by assuming that $\frac{\mu}{M}$  is a dimensionless order parameter of the PQ breaking. 
$c_i$ are $\O(1)$ coefficient. 
In essence, we are using the PQ symmetry to explain the relative smallness of $\mu$ compared to $M$, an approach commonly used to address what is known as the $\mu$-problem (see, for example, Ref.\,\cite{Weinberg:2000cr}).
For now, we will neglect the PQ breaking effect by taking $\mu \to 0$, but we will reintroduce this effect later when discussing the axion mass.

The leading term in \Eq{kahler} gives the Higgs fields negative mass squares of 
\beq
m^2_{Hu}\approx -\frac{c_u |F_Z|^2}{M^2},~~~m^2_{Hd}\approx-\frac{c_d |F_Z|^2}{M^2}.
\eeq
The magnitude of these terms can easily surpass the Hubble parameter for a sub-Planckian $M\lesssim M_{\rm pl}$ with $c_u+c_d=\O(1)(>0)$. 
The Higgs fields then fall into a configuration where
\beq
H_u^0\approx H_d^0=h_{\rm inf} \sim M, H_u^+=(H_d^-)^*\approx 0 ,
\eeq
during inflation.
Here we assumed that $H_u, H_d$ are stabilized among the negative mass terms and the higher dimensional terms suppressed by $M$ along the D-flat direction, and the field value along the D-flat direction is naturally {$\mathcal{O}(M)$}.
%$M$.
If $|F_Z/M| \ll M$ (a condition also necessary for the EFT to be perturbative), the Higgs fields can be closely approximated on the D-flat direction. This is because otherwise the D-term potentials contribute too much, $\sim g^2 M^4,g'^2M^4 (\gg |F_Z|^2),$ compared to the SUSY breaking effects.

We find that all of the particles charged under $\U(1)\times \SU(3)$ get masses of order $h_{\rm inf}\sim M$.%
\footnote{The fermion masses may not be suppressed by the small Yukawa coupling in the SM. 
The dependencies include not only $\tan\b=\vev{H_d^0}/\vev{H_u^0}$, which differs in the early Universe with a value approximated to be 1, but also the radiative corrections.
As an example, if the current smallness of the electron mass in vacuum is attributed to the large value of $\tan\b$ and radiative corrections - similar to those causing the electron's $g-2$ - the electron mass could be as large as $10^{-2}M$~\cite{Endo:2019bcj}. Higher dimensional terms may also contribute. This is an explanation for the absence of the GUT relation in the first two generation Yukawa couplings. }

\subsection{Early universe ``axion" from the MSSM}
\lac{Eaxion}
By integrating the fields of the scale $M$ with respecting SUSY, we get the EFT to be
\beq
K= K(T+T^*, \frac{Z}{M}),~~~~ {\cal L}_{\rm axion } \supset \int{d^2\theta\frac{T}{8\pi M} (c_\gamma \alpha F_{\a} F^{ \a} + c_g \alpha_s G_{a,\a} G^{a, \a})}+h.c. \, ,
\eeq
where $T$ is the NG multiplet {in the non-linear realization} of the PQ symmetry, and in the EFT the NG multiplet must appear respecting this symmetry $T\to T+i c$ with $c$ being a real number.  {Here we again neglected the explicit breaking of the PQ symmetry.}
$T$ involves the {fields in the linear realization:} $h$, $A$, and $\tl A$.
Here, $F_\a$ and $G_{a,\a}$ are the field strength superfields of $F_{\mu \nu}$ and $G_{\mu \nu}^a$ respectively.
We note that this axion in the minimal MSSM couples to photon 
with a well-known relation in the context of the QCD axion, $c_\gamma /c_g=8/3,$ which is the prediction of the MSSM. 
Therefore it is very important to study the gauge field production and GWs with this parameter relation i.e., by taking into account both $\U(1)$ and $\SU(3)$ in the analysis.
However, since this analysis is beyond the scope of the paper, in which we show that the natural UV completion of the previous studies for the GW production and that SUSY can be probed from the GWs, we do not restrict this relation further. The violation of the relation can be justified by introducing additional charged matter multiplets that couple to the Higgs doublets.\footnote{As an example, we could introduce vector lepton multiplets such as $(L', \bar L')$ and $(e', \bar e')$. $L'\AND e'$ have the same charge under the SM gauge group as those of $L$ and $e$, respectively.  
Then we consider the Yukawa couplings of $H_d L' e' \AND H_u \bar{L'} \bar{e'}$ and the Dirac mass terms $M_L\bar{L'} L', \AND M_e\bar{e'} e'$ in the superpotential. This induces additional photon coupling when $|H_{u}|=|H_{d}|\gg M_{L,e}$ during the inflation.
This Dirac mass term breaks the PQ symmetry and contributes to a potential term of the axion. Here $M_L, M_e\sim \mu$ from the naturalness.} 
The photino and gluino can acquire a SUSY breaking mass of order $|F_Z/M|$ if there presents a direct coupling to $Z$, e.g., $\int d^2\theta\frac{Z}{M} G_{a,\a}G^{a,\a}$. In addition, the term such as $\frac{Z}{M} (T+T^*)^2+h.c.$ can give a mass $\sim |F_Z/M|$ to the axino. 
Therefore, in this EFT, we expect that the superpartners of the light multiplets acquire a SUSY breaking mass of order $F_Z/M$.

Thus we have shown in the EFT below the scale of $F_Z/M$, which is larger than $H_{\rm inf}$ in order to drive the Higgs to a larger value, we get the model of \Eq{setup}. 
Here $a = \Im[T]/\sqrt{2}$ (or $A$ in linear realization) and the decay constant $ f_a\sim M$, with $h_{\rm inf}\sim M.$ \\

Lastly, we comment on the mass of $a$. The various higher dimensional terms for $H_u,H_d$ that are suppressed by $\mu$ would give a nontrivial shape of the axion potential since $H_u, H_d\sim M$ make many higher dimensional terms contribute equally. 
In the SUSY limit, the $T$ multiplet has a superpotential from \Eq{super}\footnote{There is also the axion potential from the gluon non-perturbative effect, which is negligible in the case of $H_{\rm inf}\gtrsim 100\TEV$, since the Gibbons-Hawking temperature is higher than the enhanced confinement scale (see, e.g., Ref.~\cite{Matsui:2020wfx}). On the other hand, depending on the model building, non-perturbative effects from non-abelian gauge fields may also be important.}
\beq
\D W\sim Z\mu M\(c_1 e^{i \frac{T}{M}}+c_2 e^{i 2\frac{T}{M}}+\cdots\) ,
\eeq
by approximating $H^0_u\sim H^0_d \sim M e^{i \frac{T}{2M}}$.
Then we obtain the axion potential
\beq
V(a) \simeq |F_Z\mu M| \(2|c_1|\cos(\frac{a}{f_a}+\theta_1)+2|c_2|\cos(2\frac{a}{f_a}+\theta_2) \cdots\) .
\eeq
Here $\theta_i$ are the total phases of $c_i$, $F_Z, \AND \mu$.
The axion potential height has the scale $|F_Z\mu M|$ due to the PQ breaking effect $\propto |\m|$.

{This model allows for both possibilities: the axion acting as the inflaton or as the spectator.
The axion may be the spectator if the axion potential height is much smaller than the inflaton potential scale $|F_Z|^2$, i.e., $|F_Z|\gg |\mu M|$. 
On the other hand, if $|F_Z| \sim |\mu M|$, the potential height is around the inflaton one. 
Then the axion can be the inflaton. 
Due to cancellations among different cosine terms, the curvature and the slope of the axion potential at the field point for inflation can be smaller than $|F_Z/M|$ or $|\mu|$, satisfying the slow-roll conditions. 
Our model may be seen as a UV completion of the multi-natural or ALP inflation model~\cite{Czerny:2014wza, Czerny:2014xja,Czerny:2014qqa,Higaki:2014sja, Daido:2017wwb, Daido:2017tbr,Armengaud:2019uso,Takahashi:2019qmh, Takahashi:2021tff, Takahashi:2023vhv}.}

{The axion potential height satisfies, 
\beq
\sqrt{\frac{V(a)}{f_a^2}} \sim 10^{13}\GEV \sqrt{\frac{10^{16}\GEV}{M} \frac{H_{\rm inf}}{10^{13}\GEV} \frac{\mu}{10^{11}\GEV}} ,
\eeq
where we have used $H_{\rm inf}\sim F_Z/M_{\rm pl}$. This 
can be larger than $H_{\rm inf}$, leading to fast motion of $\dot a$, which is favored to produce significant GWs.}
{
If $|c_\g|\gg |c_g|$, we can have the type of the GW spectrum as in the $\U(1)$ scenario in \Sec{setup}, similar to green lines in Fig.\,\ref{fig: GW spectrum}, while if $|c_g|\gg |c_\g| $, we have the one as in the $\SU(2)$ scenario, similar to the blue lines in Fig.\,\ref{fig: GW spectrum}. 
Interestingly, detecting the GWs can probe the SUSY with intermediate or higher scales. In addition, SUSY is a potential source of the nanohertz GWs.\footnote{An interesting topic is the relation of the intermediate scale of $\mu$ for explaining the nanohertz GWs to the usual PQ symmetry that provides the QCD axion. This is left for future work.} 
}

\subsection{An inflation model}
\lac{inf}
Although we do not specify the inflationary model, let us provide one concrete model for clarity.  
Suppose that the inflaton sector has the K\"{a}hler and superpotential of the form
\beq K_{\rm inf}=\frac{1}{2}(\tl T+\tl T^*)^2 +Z Z^*, W_{\rm inf}=Z f(\tl T),\eeq with 
a component of $\tl T$ being the inflaton field (c.f.~\cite{Kawasaki:2000yn}). $Z$ is the field in the previous discussion providing the SUSY breaking effect. 
% We do not consider $\Re \tl T$ as the inflaton to evade the $\eta$-problem ~\cite{Kawasaki:2000yn}.

One can see that the inflaton acquires an F-term potential of $V=|F_Z|^2=|f(\tl T)|^2.$  Interestingly, this class of inflation model involves the potential shape discussed in \Sec{Eaxion} where $\Im [\tl T]$ is $\Im [T]$, i.e.,  the ``axion" in the Higgs multiplets with $f$ including higher dimensional terms, while $\Re [T]$ is stabilized by higher dimensional terms in  the K\"{a}hler potential.
Given multi-cosine-terms, the inflation can be successful~\cite{Czerny:2014wza, Czerny:2014xja,Czerny:2014qqa,Higaki:2014sja, Daido:2017wwb, Daido:2017tbr,Armengaud:2019uso,Takahashi:2019qmh, Takahashi:2021tff,Takahashi:2023vhv}. 
This gives a simple possibility of driving the inflation with the inflaton incorporated in the Higgs multiplets.

%%%%%%%%%%%%%%%%%%%%%%%%%%%%%%%%%%%%%%%%%%%
\section{Conclusions and discussion}
%%%%%%%%%%%%%%%%%%%%%%%%%%%%%%%%%%%%%%%%%%%

Our work has been inspired by recent strong evidence for the stochastic gravitational wave (GW) background, as reported by NANOGrav, EPTA, PPTA, and CPTA. This breakthrough may signify the dawn of a new era in probing the early Universe. We propose a novel source of stochastic GWs within the context of the minimal supersymmetric standard model (MSSM), a theory potentially applicable at exceedingly high energy scales.

Our focus has been on the axion field in the Higgs multiplets, which exhibits dynamic properties when the Higgs field is assumed to have a large value along the D-flat direction during inflation. 
The axion motion sets off an instability in the standard $\U(1)$ and/or $\SU(3)$ gauge fields, resulting in the production of stochastic GWs.
{We have also emphasized the importance of studying the gauge field production and GW spectrum precisely with the axion coupling to both $\U(1)$ and $\SU(3)$, especially with $c_\g/c_g=8/3$, because this is the prediction of the MSSM.}

This scenario provides a straightforward UV completion of frequently analyzed models where an axion spectator/inflaton is coupled to a hidden $\U(1)$ or $\SU(N)$ gauge field absent of matter fields. The presence of nanohertz GWs could be interpreted as an indication of supersymmetry. 
% If this assertion holds true, it may also result in the production of a primordial magnetic field.
Hence, our proposed mechanism opens up exciting possibilities for studying the early Universe and unearthing new evidence of supersymmetry.
\\

So far we have discussed GWs.
{However, the amplification of the gauge field also induces the scalar perturbations, especially in the $\U(1)$ scenario, which may provide another probe to {our scenario}.}
{Moreover,} in the same setup, one can also produce a primordial magnetic field since the $\U(1)$ is SM one.\footnote{To be precise, in the context of MSSM, the ``photon" during inflation is not the conventional one, but rather a combination of $\U(1)\times \SU(2)$ if the low-energy Lagrangian does not feature $\tan\b\approx1$.} In our scenario, we do not have the Schwinger effect that prevents the growth of the magnetic field production~\cite{Turner:1987bw}. 
However, our scenario does not necessarily contradict the baryon asymmetry or bounds for the primordial magnetic field, which was discussed in the context of the Standard Model (SM)~\cite{Jimenez:2017cdr}, because the model is different. For instance, for the discussion to hold, it is important that the chirality-flipping interaction proceeds slowly enough to store a flavor-dependent lepton asymmetry~\cite{Fujita:2016igl,Kamada:2016eeb,Kamada:2016cnb}. This may not be satisfied in our MSSM, where we have an enhancement of the light Yukawa coupling by $\tan\beta$, radiative corrections, and the presence of higher dimensional terms. These can effectively wash out the primordial helical magnetic fields. 

We also mention that our concept of making matter fields heavy during inflation to create an efficient instability in the SM gauge field, thereby producing significant GWs, can be applied to more generic setups.

%%%%%%%%%%%%%%%%%%%%%%%%%%%%%%%%%%%%%%
\section*{Acknowledgments}
This work is supported by JSPS KAKENHI Grant Numbers 23KJ0088 (K.M.),   20H05851 (W.Y.),  21K20364 (W.Y.),  22K14029 (W.Y.), and 22H01215 (W.Y.). 
%%%%%%%%%%%%%%%%%%%%%%%%%%%%%%%%%%%%%%

%%%%%%%%%%%%%%% References %%%%%%%%%%%%%%%%
\bibliographystyle{apsrev4-1}
\bibliography{Ref}
%%%%%%%%%%%%%%%%%%%%%%%%%%%%%%%%%%%%%%%%%%%

\end{document}